\documentclass[12pt]{iopart}
\usepackage[utf8]{inputenc}
\usepackage{graphicx, url, hyperref, iopams, makecell}

\begin{document}
\title{pyPCG: A Python toolbox specialized for phonocardiography analysis}
\author{Kristóf Müller, Janka Hatvani, Miklós Koller, Márton Áron Goda}
\address{50/a Práter street, 1083 Budapest, Hungary}
\ead{muller.kristof@itk.ppke.hu}

\begin{abstract}
\textit{Objective}. Phonocardiography has recently gained popularity in low-cost and remote monitoring, including passive fetal heart monitoring. Development for methods which analyse phonocardiographical data try to capitalize on this opportunity, and in recent years a multitude of such algorithms and models have been published. Although there is little to no standardization in these published algorithms and multiple parts of these models have to be reimplemented on a case-by-case basis. Datasets containing heart sound recordings also lack standardization in both data storage and labeling, especially in fetal phonocardiography. \textit{Approach}. We are presenting a toolbox that can serve as a basis for a future standard framework for heart sound analysis. This toolbox contains some of the most widely used processing steps, and with these, complex analysis processes can be created. These functions can be individually tested. \textit{Main results}. Due to the interdependence of the steps, we validated the current segmentation stage using a manually labeled fetal phonocardiogram dataset comprising 50 one-minute abdominal PCG recordings, which include 6,758 S\textsubscript{1} and 6,729 S\textsubscript{2} labels. Our results were compared to other common and available segmentation methods, peak detection with the Neurokit2 library, and the Hidden Semi-Markov Model by Springer \textit{et al.} With a 30 ms tolerance our best model achieved a 97.1\% F1 score and 10.8 $\pm$ 7.9 ms mean absolute error for S\textsubscript{1} detection. \textit{Significance}. This detection accuracy outperformed all tested methods. With this a more accurate S\textsubscript{2} detection method can be created as a multi-step process. After an accurate segmentation the extracted features should be representative of the selected segments, which allows for more accurate statistics or classification models. The toolbox contains functions for both feature extraction and statistics creation which are compatible with the previous steps. pyPCG is available on \url{https://pypcg-toolbox.readthedocs.io/en/latest/}.
\end{abstract}

\noindent{\it Keywords\/}: fetal phonocardiography, heart sound detection, digital biomarkers, feature engineering, open source Python toolbox
\maketitle
\section{Introduction}
\subsection{Background}
\label{ssec:background}
Heart disease and developmental problems can be detected by observing the cardiac activity. Measurements can be made by imaging (e.g. ultrasound) or observing the electrical or acoustical side-effects of the heart cycle. Recording these side effects can be done passively and are the basis of most widespread heart monitoring methods. Certain heart diseases correspond to given anomalies in the recorded signals and by classifying the signals if they contain these signatures the underlying heart problem can be diagnosed more easily. Phonocardiogram (PCG) analysis has gained more popularity and momentum in recent years \cite{Physionet2016,PhysioNet2022} due to its easy to understand nature and as the required processing power have been met and in certain cases even surpassed by smart devices (e.g. smartphones, tablets). This is the same case with fetal heart monitoring, passive modes of monitoring have gained more popularity because of the possibility of longer-term recordings or at-home monitoring \cite{Kovacs2011-longterm}.

A healthy adult heart produces two impulses or heart sounds, called S\textsubscript{1} for the first and S\textsubscript{2} for the second sound. These are colloquially known as ``lub" and ``dub". These sounds occur at the start of systole and diastole respectively and are caused by the closure of pairs of heart valves, while the systole and diastole are quiet regions. In certain diseases exists some noise in these intervals, called murmurs, which is most likely caused by some valve dysfunction, such as incomplete closure or stiffness. These can cause extraneous stress on the heart as the system tries to correct for the lowered efficiency of the circulation \cite{heart-stress, stenosis}. Analyzing the principal heart sounds is another key approach. As mentioned above two pairs of heart valves cause the heart sounds, since they close synchronously and form a singular impulse. A small delay can be observed between the impulses generated by a valve pair and in certain cases this delay can decrease or increase (split). For example, the splitting of S\textsubscript{2} is well-known and occurs in healthy adults while breathing. However, decreased and paradoxical splitting of S\textsubscript{2} as well as wide splitting of S\textsubscript{1} can be the consequence of some underlying conductive or mechanical problem \cite{rbbb}.

\subsection{Motivation}
\label{ssec:motivation}
Further research is needed to better understand the diagnostic value of PCG, especially fetal PCG (fPCG) signals. In order to better analyze these signals (with a larger emphasis for fPCG) a specialized library of tools, which is openly available and can be easily used by researchers and medical professionals is essential. Our goal is to provide a toolbox which achieves these requirements and additionally can be extended in the future with more analysis functions if necessary. Studying the literature certain common steps and a common processing flow can be observed across solutions. This is most pronounced in classical methods with engineered features or more conventional machine learning methods (e.g. support vector machine, random forest). The basic common processing is as follows: record input, preprocessing, segmentation, feature extraction, and then finally a statistical analysis or classification model. The input signals can be in different formats, PCG signals are usually stored as \textit{.wav} files but their other parameters such as sample rate and bit depth are not standardized. Most common sampling frequencies are between 4 kHz and 44.1 kHz with bit depths of 16 bits \cite{Ephnogram,CircorDigiscope,Challenge2016Data}, however, in fetal PCG databases the sampling rate can be as low as 333 Hz with a bit depth of 8 bits \cite{FetalDataset,shiraz,indian-institute-data,fpcg-simul}. A general toolbox needs to support these highly diverse data formats. With such a framework different datasets could be used simultaneously and compared to each other, and processes developed in this framework could be cross validated with data coming from other sources.

Commonly used preprocessing methods include simple linear filtering techniques such as low-pass, high-pass, and bandpass filters. The filtering steps are utilizing the fact that the heart sounds are in a narrow frequency band, most of the energy is located between 10 Hz and 100 Hz \cite{hs-freq, NAGEL}. These filters are usually created using the Butterworth function, but the cutoff frequencies and filter orders are highly diverse. For example a 2\textsuperscript{nd} order bandpass filter is used with 25-400 Hz in \cite{Renna-UNET}, a 6\textsuperscript{th} order bandpass with 25-900 Hz in \cite{chakir2016detection}, a 10\textsuperscript{th} order lowpass with 150 Hz in \cite{papadaniil2013efficient}, and a 3\textsuperscript{rd} order highpass with 2 Hz in \cite{SQI}. Noise reduction is usually achieved with these previous filters, although more complex denoising methods can be also observed in the literature \cite{EMD-denoise, WT-denoise, wtNN-denoise}. Segmentation methods are the most diverse with simple impulse detections or envelope peak detections to the widely used and \textit{de facto} state-of-the-art hidden semi-Markov models \cite{Springer-HSMM}. Although some processes do not require segmentation, including most neural network based approaches \cite{BOZKURT2018132, Krishnan2020}.

Engineered features usually come in two forms, time domain and frequency domain signal properties \cite{Goda2016, Müller2022, Jalali2022, Imran2022, Ortiz2016, Tschannen2016}. Other popular features come from time-frequency representations such as short-time Fourier transform (STFT), Mel-frequency cepstral coefficients (MFCC), and continuous wavelet transform (CWT) \cite{chirplet-mcc, Lee2022, Jalali2022, Müller2022, Imran2022, Ortiz2016, SMiller2016}. Although less commonly used, complexity-based features such as fractal dimension estimation and Lyapunov exponent calculation show promising potential \cite{Koutsiana-WT-FD, Müller2022, Morshed2023}. Bringing these discussed methods into a standardized framework, which can be extended in the future with additional processes was our main motivation for this work. For better usability, we included some basic statistical analysis functions, although for a complete processing model, a classification algorithm should be used.

Our previous work mainly focuses on fetal PCG analysis and abnormality detection. During this research we built up a toolkit of processing methods which are not specific to fetal cases and could be used for pediatric and adult PCG processing. This toolbox is the result of further generalization of these methods.

\begin{figure}[t]
    \centering
    \includegraphics[width=0.9\textwidth]{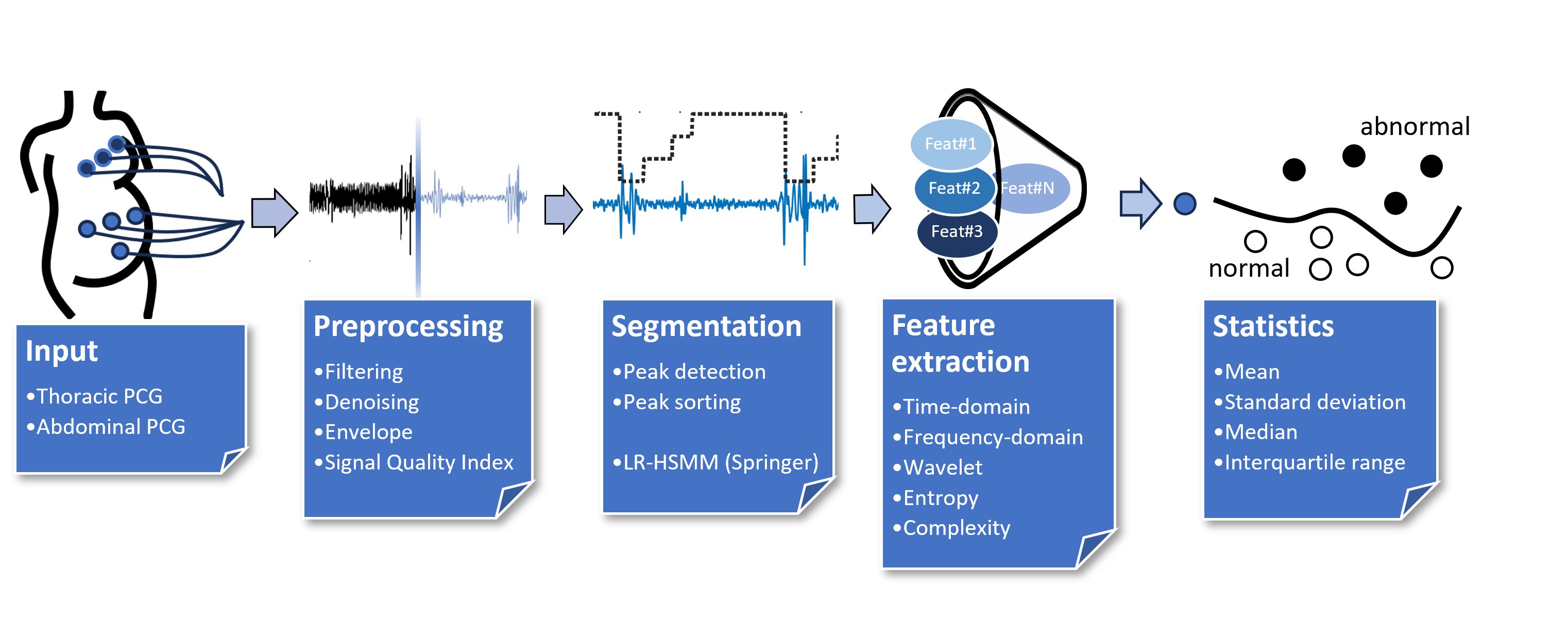}
    \caption{Major pyPCG steps and implemented functionality.}
    \label{fig:pyPCG}
\end{figure}

\section{Methods}
\subsection{Overview of the toolbox}
The implementation contains several modules to encapsulate similar functionality. These are described in the following sections in the conventional processing order:
\begin{itemize}
    \item Input and output handling (\ref{ssec:io-system})
    \item Signal quality index calculation (\ref{ssec:sqi})
    \item Preprocessing and envelope extraction (\ref{ssec:preproc})
    \item Segmentation of heart sounds (\ref{ssec:segment})
    \item Feature extraction from selected segments (\ref{ssec:features})
    \item Statistic measure calculation from the features (\ref{ssec:stats})
\end{itemize}
These blocks are illustrated in \Fref{fig:pyPCG}. Similar separate modules were not created for additional utility functions, since the code has not reached a high enough complexity for this to be an improvement. 

The central component in most processing steps is the \textit{signal object}. This abstract data object stores both the raw digital samples and the sampling rate. This way the time length of the signals can be easily acquired as well as checking the Nyquist limit for certain frequency dependent steps (e.g. filtering). The signal object also contains a log of the applied processing steps. Other abstract objects exist in the above modules, mainly for packing similar functionality into a repeatable system, and will be described in the appropriate sections.

\subsection{Input output system}
\label{ssec:io-system}
PCG recordings can be saved in several formats, which vary from dataset to dataset. Most common formats, namely \textit{.wav}, raw binary, and Matlab \textit{.mat} are supported as inputs. In the case of the \textit{.mat} format the file must contain two variables: \textit{sig} for the sample data and \textit{fs} for the sampling frequency. Similarly, when using the raw binary format, the sampling frequency must be specified manually. Less common, but important in our case, is the Fetaphon format \cite{Kovacs2011}, which is a binary file sampled at 333 Hz.  Our software reads these files and corrects the shift errors specific to this format.

The input system also has the functionality to read in files containing time labels for each heart sound. This was implemented for validation and training of the segmentation methods. The annotation files are stored as character separated values (\textit{.csv}) with a semicolon (\textit{;}) delimiter. There are two columns in these files, ``\textit{Location}" and ``\textit{Value}", where the \textit{Location} is the time of the label in seconds, and the \textit{Value} is either ``S\textsubscript{1}" or ``S\textsubscript{2}", other values are ignored.

\subsection{Signal quality}
\label{ssec:sqi}
The implemented Signal Quality Indices (SQI) are based on the work of Tang \textit{et al.} \cite{SQI}. These measures were selected based on their efficacy in differentiating between acceptable and lower quality signals, based on the results presented in the same article.

For the first measure, we selected the kurtosis of the signal values, defined as
\begin{equation}
    K_x=\frac{E \left[ x^4 \right]}{E \left[ x^2 \right]^2},
\end{equation}
where $x$ is the raw signal value, and $E[\cdot]$ is the expected value operator. Other measures related to time domain information used the envelope of the signal, $H(n)$, for which the calculation can be seen in the following \ref{ssec:preproc} Preprocessing section. These SQIs were the standard deviation (STD), the maximum value of the autocorrelation function within a specified interval, and the sample entropy (SampEn) of the envelope. The normalized autocorrelation function of the envelope ($H(n)$) is defined as
\begin{equation}
    R(l)=\sum_{n=-\infty}^{+\infty}H(n)H(n-l) \left( \sum_{n=-\infty}^{+\infty}H(n)^2 \right)^{-1},
\end{equation}
where $n$ is the discrete sampling time and $l$ the \textit{lag-time}.

The \textit{SampEn} of a signal is used to measure the complexity of the input, with a higher value representing a higher amount of noise. To define SampEn assume a signal $x(n)=\{x_1,x_2,x_3,...,x_N\}$, a template from the signal with length $m$ $x_m(n)=\{x_n,x_{n+1},x_{n+2},...,x_{n+m-1}\}$, and a distance function $d[x_m(i),x_m(j)]$ where $i\neq j$. Then define $A$, based on a tolerance parameter $r$, as the number of template pairs in the signal having $d[x_{m+1}(i),x_{m+1}(j)]<r$, and similarly $B$, as the number of template pairs in the signal having $d[x_m(i),x_m(j)]<r$. Then the definition for sample entropy is 
\begin{equation}
    SampEn = -\ln \frac{A}{B},
\end{equation}
Values for $m$ and $r$ are usually chosen as 2 and 0.2$\times$standard deviation of the data, respectively. The Python implementation was adapted from \cite{SQI} and \cite{SampEn}.

The last measure we selected was the degree of periodicity based on the cycle frequency spectrum \cite{CyclicSpectrum}. PCG signals were previously described as quasiperiodic \cite{Kumar_2011} which opens the possibility for periodicity measurement in the cycle frequency domain. First the time-varying autocorrelation ($R_x(t,\tau)$) is defined as
\begin{equation}
    R_x(t,\tau) = \lim_{N \to +\infty} \frac1{2N+1}
    \sum_{n=-N}^{N}x \left(t+\frac{\tau}{2}+nT \right)x^*\left(t-\frac{\tau}{2}+nT \right),
\end{equation}
where $T$ is the cycle duration of the heart sound signal $x(t)$, and $x^*(t)$ is its complex conjugate. The second time parameter $\tau$, is known as the \textit{lag-time}.
Since this is a periodic function, it can be written in using its Fourier series representation as
\begin{equation}
    R_x(t,\tau)=\sum_{\alpha=-\infty}^{+\infty}R_x(\alpha,\tau)e^{i2\pi\alpha t},
\end{equation}
where $\alpha$ is called the cycle frequency.
The Fourier series coefficients are calculated as
\begin{equation}
    R_x(\alpha,\tau) = \left\langle x(t+\tau/2)x^*(t-\tau/2)e^{-i2\pi\alpha t} \right\rangle_t,
\end{equation}
where the $\left\langle\cdot\right\rangle_t$ operator is used for the time average. This function $R_x(\alpha,\tau)$ is called the cyclic correlation function.
Transforming the cyclic correlation to frequency domain gives us the cyclic spectral density written as
\begin{equation}
    S_x(\alpha,f) = \int_{-\infty}^{+\infty} R_x(\alpha,\tau) e^{-i2\pi\tau f}d\tau.
\end{equation}
Integrating over the frequencies gives us the cycle frequency spectral density (CFSD),
\begin{equation}
    \gamma_x(\alpha) = \int_{-\infty}^{+\infty} \left| S_x(\alpha,f) \right| df.
\end{equation}
From the CFSD the degree of periodicity is defined as
\begin{equation}
    dp_x =\frac{\max (\gamma_x(\alpha))}{\mathrm{median}(\gamma_x(\alpha))},
\end{equation}
where the $\max(\cdot)$ and $\mathrm{median}(\cdot)$ operators give the maximum and the median values of their arguments for all possible $\alpha$ inputs respectively.

\subsection{Preprocessing}
\label{ssec:preproc}
The toolbox also includes several steps that operate on the entire signal. These steps are referred to as \textit{preprocessing} functions in our definition. Their input is a signal object with supplementary parameters, and the output is another signal object with the specified operation applied. One of the more straightforward operations is the slicing of the input into shorter length signals with a preset constant length and the option of overlap. Another simple operation is resampling the signal, which is a wrapper on the SciPy \textit{resample\_poly} function to make it compatible with the signal object. Classical filtering is also implemented using the Butterworth filter design, with settable filter order, frequency, and type (e.g. low-pass, high-pass).

\begin{figure}
    \centering
    \includegraphics[width=0.9\columnwidth]{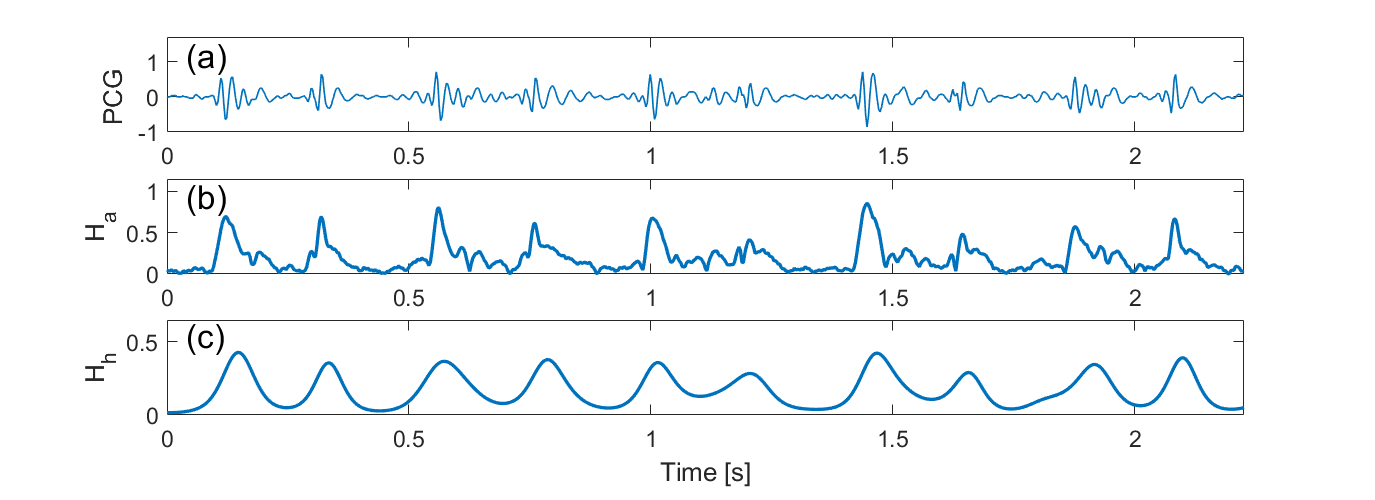}
    \caption{An example PCG signal (a), its Hilbert envelope (b), and homomorphic envelope (c).}
    \label{fig:henv}
\end{figure}

Two types of envelope calculations are included, visual examples for these can be seen in \Fref{fig:henv}. The simple version $H_\mathrm{a}(t)$ calculates the envelope with the Hilbert transformation, which transform is defined as
\begin{equation}
    \hat{x}(t)=\frac1\pi \int_{-\infty}^{+\infty}\frac{x(k)}{t-k}dk,
\end{equation}
where $x(t)$ is the original signal and $\hat{x}(t)$ is its Hilbert transform pair. With it the complex valued analytical signal $x_\mathrm{a}(t)$ is defined as
\begin{equation}
    x_\mathrm{a}(t) = x(t)+i \; \hat{x}(t)
\end{equation}
The analytical signal can be rewritten in an exponential form as
\begin{equation}
    x_\mathrm{a}(t) = A(t)\rme^{\rmi\pi\phi(t)},
\end{equation}
where $\phi(t)$ is the instantaneous phase, and $A(t)$ is the instantaneous amplitude or the Hilbert envelope ($H_\mathrm{a}(t) = A(t)$) of the signal.
Another popular envelope is the homomorphic envelope, which is achieved with the homomorphic filtering of the Hilbert envelope \cite{Mubarak2018-uc}. This special filter consists of a nonlinear function (usually a logarithm), a linear filter, and the inverse of the nonlinear function (e.g. exponentiation). The mathematical description of homomorphic filtering is
\begin{equation}
    H_\mathrm{h}(t) = \mathrm{exp} \left(LPF \left(\:\log \left( H_\mathrm{a} \left(t \right) \right) \right) \right),
\end{equation}
where $LPF$ is a low-pass filter. We used the above definition with a first order Butterworth low-pass filter at 8 Hz cutoff frequency as detailed in \cite{Springer-HSMM}. A visual example for the homomorphic envelope can be seen in \Fref{fig:henv}-c.

Multiple denoising methods were also implemented in our toolbox, mainly based on the work of \cite{EMD-denoise} and \cite{WT-denoise}. They used wavelet decomposition and empirical mode decomposition (EMD) to create a multi-scale denoising method. The implemented functions apply a soft threshold on all decomposition levels with an automatically calculated threshold value, but it is also possible to set this threshold manually. Another method uses the Savitzky-Golay filter \cite{savgol} to smooth out the decomposition levels, the default behavior uses a window size of 10 samples and a degree of 3. 

The above functions take inputs and give outputs in the same formats, so they can be chained together. An abstract object called a \textit{processing pipeline} was implemented to facilitate this. It can contain preconfigured functions from this module and calls these functions one-by-one. Once the pipeline object is created, it can be used multiple times with different inputs.

\subsection{Segmentation}
\label{ssec:segment}
PCG signals are characterised by two primary descriptors, namely the S\textsubscript{1} and S\textsubscript{2} sounds. The toolbox supports two principal types of segmentation techniques for their localization, one based on envelope peak detection and the other on the Hidden Semi-Markov Model (HSMM).

Peak detection can be achieved with a simple local maxima finding function, a fast and relatively accurate method. However, it requires the fine-tuning of several parameters, such as a threshold value, the minimal distance between detections, and others \cite{2020SciPy-NMeth}. Additionally, this naive method is not robust against noise, so multiple peaks will be missed and reported as false negatives.

For a more robust detection method we developed an adaptive method based on the peak detection of the Pan-Tompkins algorithm \cite{Pan-tompkins}. First, all local maxima were extracted which gave us the first candidates. Subsequently, each candidate was subjected to a sequential filtration process, whereby any significant drop in value was detected between the selected and subsequent candidate points. This value drop was set to be proportional to the value at the selected point, with a proportionality factor of 50\%. This way we could detect the prominent peaks in the input envelope, which corresponded to both of the S\textsubscript{1} and S\textsubscript{2} sounds.

To differentiate between S\textsubscript{1} and S\textsubscript{2} sounds a simple method was developed, where the delays between each detection point were measured. This was based on the biological fact that at normal heart rates (below 80 BPM) the systole interval is shorter than the diastole interval \cite{Heartrate-1}. Although, for fetal cases some evidence suggests that this is not the case \cite{Heartrate-fetal-2}. However in \cite{fpcg-simul} the time difference between S\textsubscript{1} and S\textsubscript{2} events were shorter than S\textsubscript{2}-S\textsubscript{1} intervals, and in we noticed the same pattern in our data (see \Tref{tab:data-sd}). Thus in our peak sorting method, if the measured delay was shorter than the previous one, it was considered a systole interval corresponding to an S\textsubscript{1} at its beginning and an S\textsubscript{2} at its end. By the same logic, the other intervals corresponded to diastole, which meant an S\textsubscript{2}-S\textsubscript{1} pair in the detections. Since feature calculation steps require event start and end times, the detected peaks need to be converted to these time-pairs. This is achieved with observing the envelope value at the detected peak and measuring the time points before and after the detection when the envelope value drops below a set percentage of the peak.

HSMM with logistic regression (LR-HSMM) was described by Springer \textit{et al.} \cite{Springer-HSMM}, and a MATLAB implementation \cite{Springer-HSMM-impl} was made public by the authors on PhysioNet \cite{PhysioNet}. This method requires training on sample data, in contrast to the previous algorithms. The problem with such a method is that for successful training a large enough pre-labeled dataset is needed. This training is also time-consuming, however, the trained model can be easily reused. The HSMM process as described by Springer \textit{et al.} calculates different envelograms of the signal and trains a logistic regression model to give the probability of the heart cycle being in a certain state (S\textsubscript{1}, systole, S\textsubscript{2}, diastole). Then based on these observations, the heart cycle state transitions, and the time length of the current state, an extended Viterbi algorithm \cite{Springer-HSMM} estimates the most probable state sequence. This state sequence gives us the segmentation of the PCG signal. The open MATLAB implementation was reimplemented in Python using the \textit{hsmmlearn} package \cite{hsmmlearn}, which is a direct port of the R \textit{hsmm} package \cite{hsmm-r}. Slight modifications were made from the original implementation by Springer to better work with our data and use cases. Most notably a different set of parameters were used for the included filters, heart rate estimation, systole length estimation, and expected heart sound lengths. This implementation will be referred to as \textit{pyHSMM} in the rest of this article. In order to enhance the reusability of trained pyHSMM instances, a serialization and unserialization methodology has been developed. This way pretrained models can be loaded from files.

\begin{table}[t]
\caption{Implemented features in the current version of pyPCG}
\fontsize{8pt}{10.5pt}
\selectfont
\begin{tabular}{@{}lll}
\br
Feature name & Definition & Reference \cr \mr
\textbf{\textit{Time domain}}   & & \cr
1 Time-delta    & Difference in time between the start and end points [ms] & \makecell[tl]{\cite{Goda2016, Müller2022, Jalali2022}\\\cite{Imran2022, Ortiz2016, Esmail2019}} \cr
2 Onset time    & Time duration from start to envelope maximum [ms] & \cr
3 Exit time     & Time duration from envelope maximum to end [ms] & \cr
4 Peak spread   & \makecell[tl]{Time length of interval around the envelope maximum\\with a proportional area under curve to the complete area [ms]} & \cr
5 Peak width    & \makecell[tl]{Time length between points before and after\\the envelope maximum with a proportion of the maximum value [ms]} & \cr
6 Peak centroid & \makecell[tl]{Time from start to center of gravity, given by the point\\where the area under curve is half of the complete area [ms]} & \cr
7 Zero-crossing rate & Number of value sign changes in a unit of time [1/ms] &\cite{Müller2022, Esmail2019, Yadav2020} \cr \mr
\textbf{\textit{Frequency domain}} & & \cr
8 Maximal frequency & Frequency value with the largest amplitude [Hz] & \cr
9 Spectral spread   & \makecell[tl]{Frequency range of interval around the spectrum maximum\\with a proportional area under curve to the complete area [Hz]} & \cr
10 Spectral width    & \makecell[tl]{Frequency range between points before and after the\\spectrum maximum with a proportion of the maximum value [Hz]} & \cr
11 Spectral centroid & \makecell[tl]{Frequency of the center of gravity, given by the point\\where the area under curve is half of the complete area [Hz]} & \cite{Yadav2020, Zabihi2016} \cr \mr
\textbf{\textit{Wavelet based}} & & \cr
12 Scale of maximum coefficient & Pseudofrequency of the largest absolute CWT coefficient [n.d.] & \cr
13 Time of maximum coefficient  & Time of the largest absolute CWT coefficient [ms] & \cr
14 CWT peak distance            & \makecell[tl]{Euclidean distance between the two largest absolute local maxima\\of CWT coefficients. (Zero if only one maximum exists.) [n.d.]} & \cr
15 DWT intensity                & Root mean square of specified DWT detail components [n.d.] & \cr
16 DWT entropy                  & \makecell[tl]{Root of second order Shannon entropy of specified\\DWT detail components [n.d.]} & \makecell[tl]{\cite{Müller2022,Jalali2022}\\\cite{Zabihi2016,Li2019}} \cr \mr
\textbf{\textit{Complexity based}} & & \cr
17 Katz fractal dimension & \makecell[tl]{Ratio of the logarithm of the time length of the signal and\\the sum of this logarithm with the logarithm of the ratio between\\the ``diameter" of the signal and its curve length. [n.d.]} & \cite{Koutsiana-WT-FD, Morshed2023} \cr
18 Lyapunov exponent      & \makecell[tl]{In a dynamical system, the rate of separation of\\two infinitesimally close trajectories [n.d.]} & \cite{Müller2022, Karar2017} \cr
\br
\end{tabular}
\label{tab:feature-defs}
\end{table}

\subsection{Feature extraction}
\label{ssec:features}
A number of signal features were implemented in accordance with the general trends observed with classical methods. These engineered features can be grouped based on the domain they are calculated from (see Section \ref{sssec:timedomain} for the time, \ref{sssec:frequencydomain} for the frequency, \ref{sssec:waveletdoamin} for various wavelet domain, and \ref{sssec:complexity} for complexity-based features). For a complete summary of the implemented features, see \Tref{tab:feature-defs}. These features were chosen to facilitate the detection of abnormalities in the PCG signal corresponding with heart diseases, as detailed in Section \ref{ssec:background}. In our implementation, the feature calculation functions require as input the signal and a time interval where the calculation should occur.

From the segmentation methods, we get the boundaries of the S\textsubscript{1} and S\textsubscript{2} sounds. These can be used to acquire the systole and diastole regions, as well as the whole heart cycle, by changing what we consider the beginning or the end of a given event. For example entire heart cycles can be passed by setting S\textsubscript{1} beginning times as the start, and shifted  S\textsubscript{1} beginning times as the end markers of the events.
\subsubsection{Time domain features:}
\label{sssec:timedomain}
\begin{figure}
    \centering
    \includegraphics[width=0.45\columnwidth]{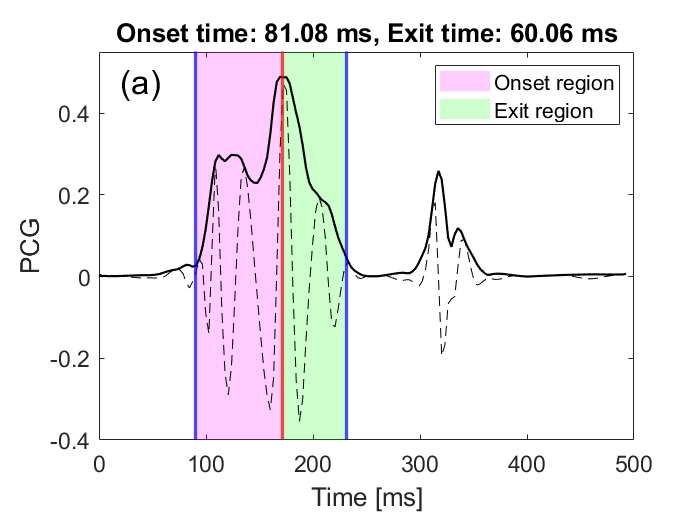}
    \includegraphics[width=0.45\columnwidth]{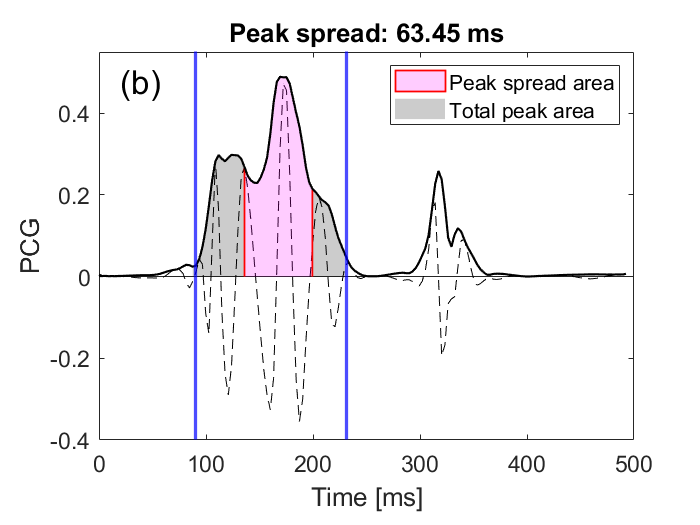}
    \includegraphics[width=0.45\columnwidth]{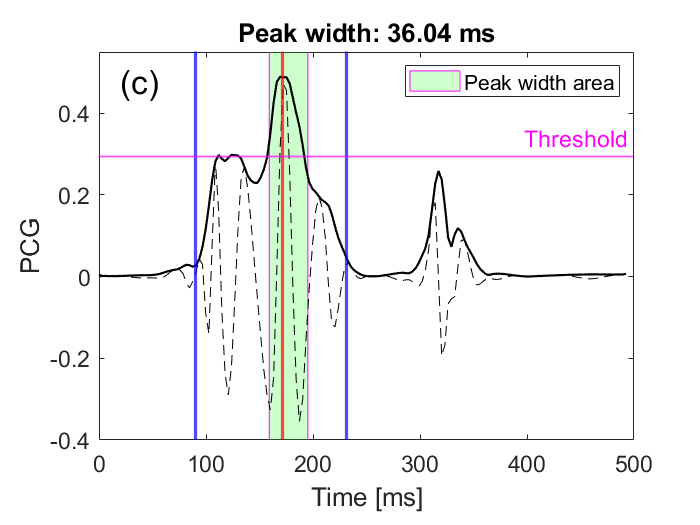}
    \includegraphics[width=0.45\columnwidth]{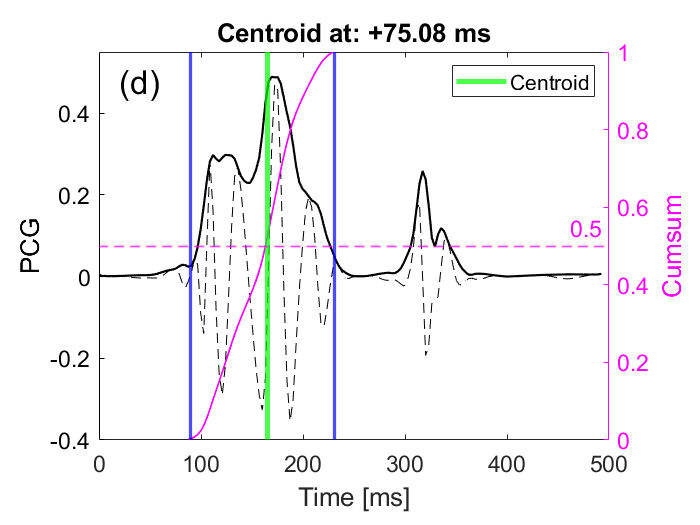}
    \caption{Visualisation of time domain features: (a) ramp time, (b) peak spread, (c) peak width, (d) peak centroid. On each figure the dashed line shows the PCG signal, the solid black line the Hilbert envelope, the blue lines mark the region of interest (S\textsubscript{1}), and the red lines mark the maximum location of the envelope. A ratio factor of 0.6 was used in peak width and peak spread. The time-delta of this region is 141.14 ms and the zero-cross rate is 0.24.}
    \label{fig:time-domain}
\end{figure}

The simplest feature we can calculate is the difference between the event start and end times. This is called the \textit{time-delta} feature. Another time difference we can calculate is the \textit{onset time} and \textit{exit time} (collectively, called as \textit{ramp time}). In our definition \textit{onset time} is the delay between the start and the peak, and \textit{exit time} is the delay between the peak and the end marker. For visual representation in the case of an S\textsubscript{1} sound see \Fref{fig:time-domain}-a. We also defined the \textit{peak spread} and \textit{peak width} features, these can give similar results, however they are defined in different ways. Peak spread is calculated by summing the envelope values between the event boundaries, and selecting an interval around the local maximum with an area that represents a given percentage of the total area. The time length of this new interval is the peak spread. The peak width is the difference between the preceding and succeeding time locations where the value is at a given proportion of the peak value.
These are illustrated in \Fref{fig:time-domain}-b and \Fref{fig:time-domain}-c respectively. Another unique time domain feature we introduced is the \textit{peak centroid}, defined as the ``center of gravity" of the selected interval. This can be readily determined through the application of a cumulative sum to the envelope, identifying the first value that exceeds the half of the total sum of the envelope. This is illustrated in \Fref{fig:time-domain}-d. The last current feature considering the time domain is the zero-crossing rate, defined as the number of sign changes in the signal in a unit of time. This was implemented to replicate the behavior of the \textit{zerocrossrate} MATLAB function \cite{matlab-zcr}.
\subsubsection{Frequency domain features:}
\label{sssec:frequencydomain}
\begin{figure}
    \centering
    \includegraphics[width=0.45\columnwidth]{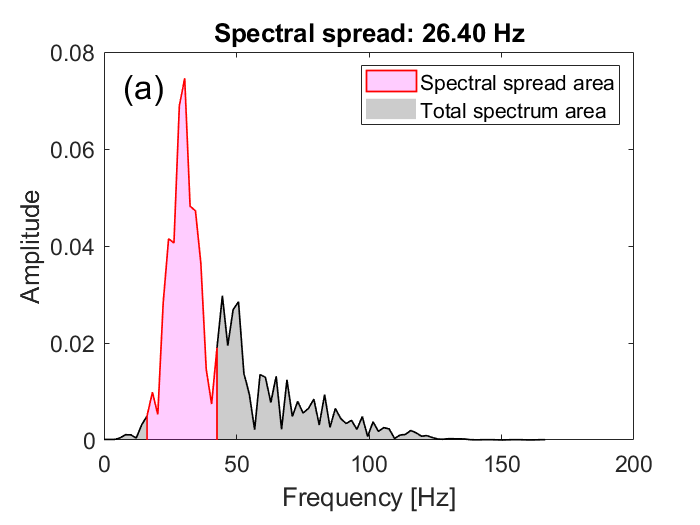}
    \includegraphics[width=0.45\columnwidth]{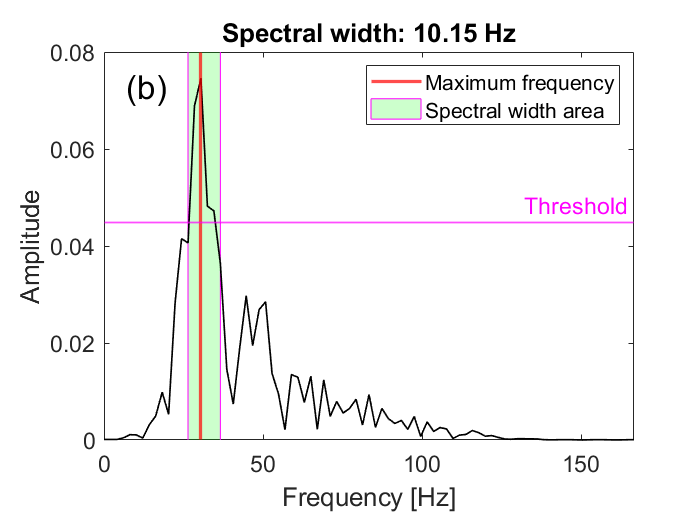}
    \includegraphics[width=0.45\columnwidth]{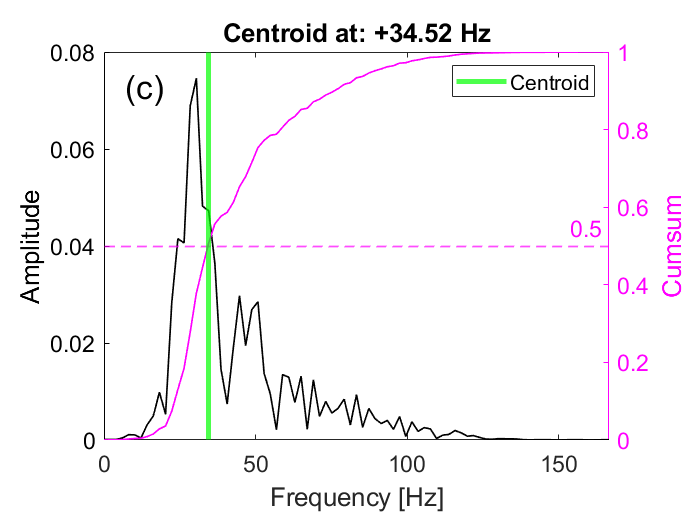}
    \caption{Visualisation of frequency domain features: (a) spectral spread, (b) spectral width, (c) spectral centroid. On each figure the solid black line marks the Fourier spectrum. A ratio factor of 0.6 was used in spectral width and spectral spread. The maximal frequency is 30.46 Hz.}
    \label{fig:freq-domain}
\end{figure}

Frequency domain information has been suggested previously to be advantageous in detecting certain abnormalities. In our previous work \cite{fetal-split} we showed that the Fourier spectrum can be used with high accuracy to detect split S\textsubscript{1} in fetal PCG. The current toolbox supports calculating multiple features in the frequency domain. Such as the \textit{maximal frequency}, which is the frequency of the given component with the highest energy. Similarly to the time domain features, we introduced \textit{spectral spread}, \textit{spectral width}, and \textit{spectral centroid}. These have similar definitions to their time domain counterparts, however, the time differences or intervals are instead defined as frequency differences and intervals. Visual examples for these can be seen in \Fref{fig:freq-domain}.

\subsubsection{Wavelet based features:}
\label{sssec:waveletdoamin}
\begin{figure}
    \centering
    \includegraphics[width=0.9\columnwidth]{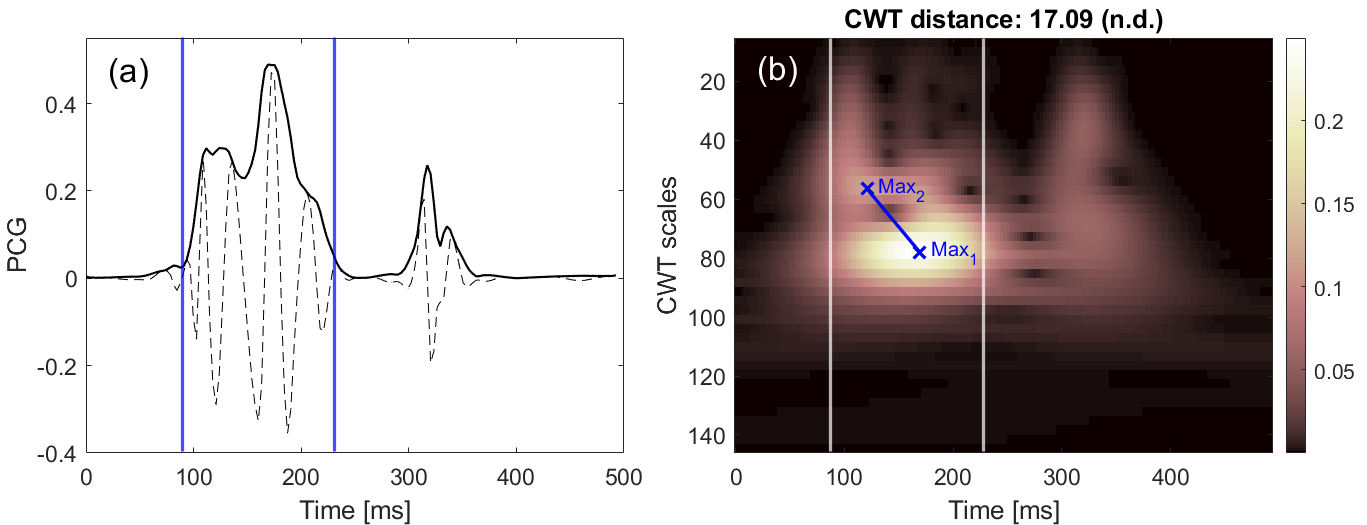}
    \caption{Visualisation of CWT distance feature. Time domain signal (a) and its CWT (b). The dashed line means the PCG signal and the solid black line is its Hilbert envelope. The vertical blue and white lines mark the region of interest (S\textsubscript{1}). Local maxima marked as Max\textsubscript{1} and Max\textsubscript{2}}
    \label{fig:cwt-dist}
\end{figure}

Wavelet transformation and decomposition techniques are applied extensively in the field of  biomedical signal processing, and wavelet techniques were demonstrated in multiple PCG analysis publications \cite{Koutsiana-WT-FD, Kovacs2011}. It is important to differentiate between the continuous wavelet transform (CWT) and discrete wavelet transform (DWT) often referred to as wavelet decomposition \cite{Kovacs2011, DWT-other}. CWT is usually defined as
\begin{equation}
    W(a,b) = \frac1{\sqrt a} \int_{-\infty}^{+\infty}x(t) \psi^* \left( \frac{t-b}a \right)dt,
    \label{eq:WT}
\end{equation}
where $\psi(t)$ is the mother wavelet, a finite basis function, and $\psi^*(t)$ is its complex conjugate. Scaling parameter $a$ is analogous to the frequency of the Fourier transform, whereas translation parameter $b$ determines the position of the time window where the mother wavelet is applied \cite{Kovacs2011}. The CWT can be obtained for any $a$-$b$ pairs. The visual representation of this function is called a scalogram. The toolbox currently contains two engineered CWT-based features. The first one is the \textit{maximum coefficient}, returning both the position in time and scaling factor of the maximum coefficient, based on the set parameters. The other feature, known as \textit{CWT peak distance}, was inspired by visual inspection of the CWT scalogram across multiple signals. We noticed that in certain cases for the heart sounds the two valve components can be observed as two separate peaks. To quantitatively measure this phenomenon we measure the time and frequency distances between the two local maxima in the CWT representation of the given section. This is illustrated in \Fref{fig:cwt-dist}.

\begin{figure}
    \centering
    \includegraphics[width=0.9\columnwidth]{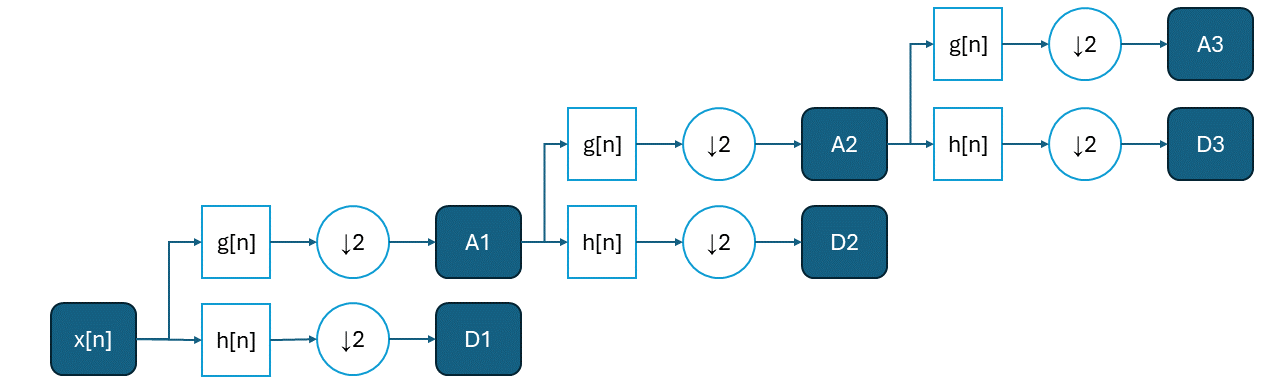}
    \caption{A model for discrete wavelet transform with three levels of decomposition: D1,D2,D3 as the detail components; A1,A2,A3 as the approximation components; $\downarrow$2 as the downsampling stage.}
    \label{fig:dwt-3}
\end{figure}

\begin{figure}
    \centering
    \includegraphics[width=0.9\linewidth]{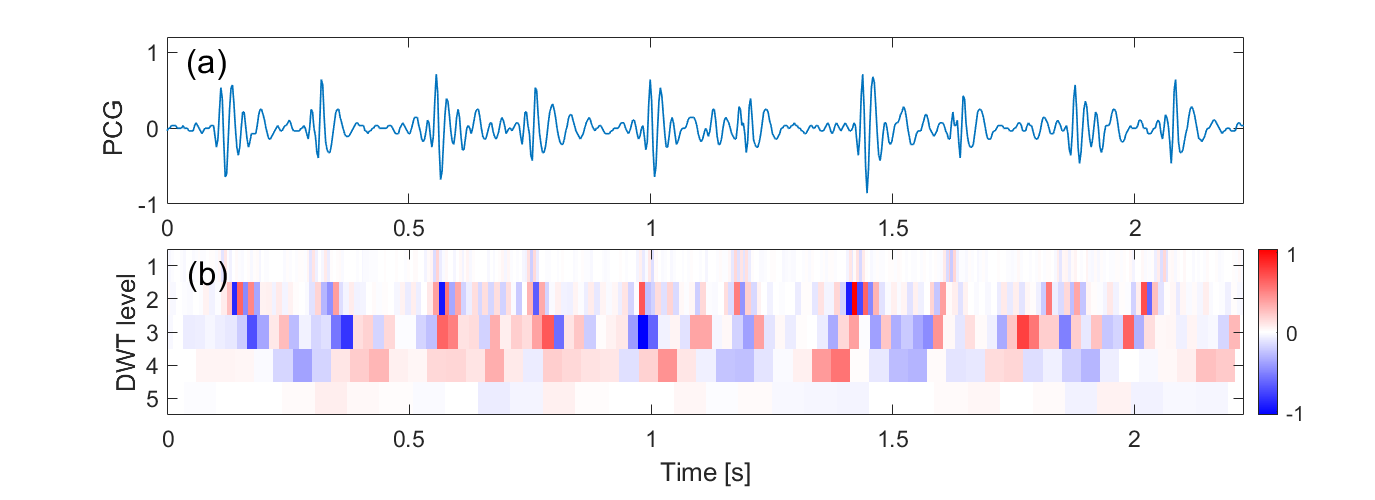}
    \caption{DWT scalogram of an example signal (a), 5 level DWT with the \textit{db6} wavelet family (b).}
    \label{fig:dwt-scalogram}
\end{figure}

DWT on the other hand allows the application of (\ref{eq:WT}) only for discrete $a$-$b$ pairs \cite{DWT-def}. In practice it is defined as a cascading process with specially defined filter pairs and downsampling steps. These filter pairs are called quadrature mirrors, one is a low-pass and the other is a high-pass filter, denoted with impulse responses of $g[n]$ and $h[n]$, respectively. Convolving the input signal $x[n]$ with one of these filters gives us the filtered signal
\begin{equation}
    y[n] = (x * h)[n] = \sum_{k=-\infty}^{+\infty}x[k]h[n-k],
    \label{eq:DWTconv}
\end{equation}
where the $*$ operator marks the convolution.
Since half of the frequencies have been removed, the filtered signal can be downsampled with a factor of two, which can be calculated as
\begin{equation}
    y[n]_{\mathrm{high}} = \sum_{k=-\infty}^{+\infty}x[k]h[2n-k],
    \label{eq:DWTds-1}
\end{equation}
\begin{equation}
    y[n]_{\mathrm{low}} = \sum_{k=-\infty}^{+\infty}x[k]g[2n-k].
    \label{eq:DWTds-2}
\end{equation}
These two results $y[n]_{\mathrm{high}}$ and $y[n]_{\mathrm{low}}$ are known as the \textit{detail} and \textit{approximation} coefficients respectively. By further processing the approximation coefficients with another filtering and downsampling stage, another level of decomposition can be achieved, for example a three-level decomposition cascade can be seen in \Fref{fig:dwt-3}. It is noteworthy that the convolution (\ref{eq:DWTconv}) and the downsampling (\ref{eq:DWTds-1}, \ref{eq:DWTds-2}) of the DWT are analogous to the steps of the CWT in (\ref{eq:WT}). A visual example for a discrete wavelet decomposition can be seen in \Fref{fig:dwt-scalogram}.

In its current version, the toolbox supports two DWT-based features, these are called \textit{DWT intensity} and \textit{DWT entropy}. In both cases, the decomposition depth, the wavelet family, and the given decomposition level can be selected. DWT intensity is calculated as the root mean square of the values in the selected interval. This measure ($I_{\mathrm{DWT}}$) for a segment with length $N$ is
\begin{equation}
    I_{\mathrm{DWT}} = \sqrt{\frac1N\sum_{n=1}^{N}x(n)^2}.
\end{equation}
For DWT entropy ($S_{\mathrm{DWT}}$) the following definition is used for Shannon entropy \cite{shannon-entropy}
\begin{equation}
    S_{\mathrm{DWT}} = -\sum_{n=1}^{N}x(n)^2\log_2{x(n)^2}.
\end{equation}

\subsubsection{Complexity based features:}
\label{sssec:complexity}
Several publications have described the PCG generation process as a non-linear system \cite{PCG-complexity, PCG-nonlin}. To describe non-linear relations in the heart from the PCG signal, several non-linear or complexity-based features need to be introduced. One such measure we introduced is the Katz fractal dimension (KFD). This was motivated by the work by Koutsiana \textit{et al.} \cite{Koutsiana-WT-FD}. An important sidenote is that most popular Python libraries where KFD is included use an incorrect implementation which gives different results to those in the original publication by Katz \cite{KFD}. To circumvent this we reimplemented the algorithm based on the original BASIC implementation included in the original article.

Another complexity measure is the Lyapunov exponent \cite{PCG-Lyapunov}. By treating the heart sound as the output of a chaotic dynamical system, we can estimate the complexity of such a system. The Lyapunov exponent gives the separation of two infinitesimally close trajectories in the phase space of the dynamical system. However, in the case of PCG signals, this phase space is unknown and needs to be reconstructed. This can be achieved by creating an embedding space based on time-delayed versions of the signal \cite{phasespace}.
If we treat our signal $x(n)$ as a vector, then the $m$ dimensional phase space reconstruction matrix \cite{phasespace-math} $X(n)$ is
\begin{equation}
    X(n) = [x(n),x(n+\tau),x(n+2\tau),...,x(n+(m-1)\tau)],
\end{equation}
where $\tau$ is the time delay parameter which needs to be carefully selected as well as the embedding dimension. In this phase space, we can calculate the maximum Lyapunov exponent, which will be the result of this feature calculation function. It is important to note, that implementation in Python libraries and the MATLAB \textit{lyapunovExponent} function does not give the same result. This may be due to different parametrizations or internal implementations.

\subsection{Statistics}
\label{ssec:stats}
The above features are calculated for each input segment provided and they return a given value for each segment. These functions can be joined into \textit{feature groups}, which is an abstract object containing the types and parameters of each feature. This is useful if the same features need to be calculated for multiple types of segments, this way a feature group can reduce code repetition. The result of running a feature group is a data structure with key-value pairs, where the keys are the names of the features (previously configured) and the corresponding values are the arrays containing the calculated feature for each segment. These lists can be directly used in basic statistic calculations. We provided some of the most commonly used functions (mean, standard deviation, interquartile range, etc.), and a way to export such values as Pandas DataFrames and Excel spreadsheets.

\section{Validation}
The segmentation methods in the current version of the toolbox were validated with a fetal PCG database consisting of 50 manually labeled 1-minute-long recordings from multiple subjects. The heart rate distributions are shown in \Tref{tab:data-hr} with the corresponding S\textsubscript{1}-S\textsubscript{2} and S\textsubscript{2}-S\textsubscript{1} durations shown in \Tref{tab:data-sd} after outlier removal. This dataset contains 6758 S\textsubscript{1} and 6729 S\textsubscript{2} labels.

The toolbox is designed in a manner that allows each component to be evaluated separately from the others. However, in a realistic situation, where a complete process is concerned, the given stages are highly interdependent. This dependency is most pronounced in the case of the segmentation stage, since the preceding preprocessing contains only well supported processes, and the subsequent feature extraction stage is highly sensitive to the accuracy of the segments. Thus, it was decided to focus on the validation and accuracy measurement of the segmentation methods.

\subsection{Tested segmentation methods}
The main focus of the validation was to evaluate the accuracy of detecting the first heart sound, however, in certain cases the S\textsubscript{2} detection accuracy was also measured. The tested methods were the following:
\begin{itemize}
    \item local maximum detection on the envelope of the signal (SciPy implementation \cite{2020SciPy-NMeth}), with preset minimum distances between peaks (labeled as ``\textit{naive}")
    \item our adaptive local maximum detection on the homomorphic envelope, with a basic heart sound differentiation method (labeled as ``\textit{adaptive}")
    \item local maximum detection with the NeuroKit2 implementation \cite{Neurokit} on the homomorphic envelope, with no heart sound differentiation (labeled as ``\textit{Neurokit}")
    \item Hidden Semi-Markov Model based on \cite{Springer-HSMM} (labeled as ``\textit{HSMM}")
    \item our implementation of the HSMM model with logistic regression (labeled as ``\textit{pyHSMM}")
\end{itemize}
\begin{table}[t]
\setlength{\tabcolsep}{15pt}
\caption{Heart rate distribution of the testing data}
\begin{center}
\begin{minipage}{300pt}
\begin{tabular}{@{}lllll}
\br
No. records & MEAN & STD & MIN & MAX \cr \mr
50 & 138.2 & 3.5 & 133 & 144 \cr
\br
\end{tabular}
\fontsize{8pt}{10.5pt}
\selectfont
\textsuperscript{*}The mean, standard deviation (STD), minimum (MIN), and maximum (MAX) heart rates reported in the dataset, measured in beats-per-minute (BPM).
\label{tab:data-hr}
\end{minipage}
\end{center}
\end{table}
\begin{table}[t]
\setlength{\tabcolsep}{17pt}
\caption{S\textsubscript{1}-S\textsubscript{2} and S\textsubscript{2}-S\textsubscript{1} duration distribution in the testing data}
\begin{center}
\begin{minipage}{300pt}
\begin{tabular}{@{}lllll}
\br
Section & MEAN & STD & MIN & MAX \cr \mr
S\textsubscript{1}-S\textsubscript{2} & 181 & 21 & 69 & 270 \cr
S\textsubscript{2}-S\textsubscript{1} & 257 & 24 & 183 & 333 \cr
\br
\end{tabular}
\fontsize{8pt}{10.5pt}
\selectfont
\textsuperscript{*}The mean, standard deviation (STD), minimum (MIN), and maximum (MAX) durations reported in the dataset after outlier removal, measured in ms.
\label{tab:data-sd}
\end{minipage}
\end{center}
\end{table}
\subsection{Accuracy measures}
Detection accuracy is a non-trivial task, since a small delay between the manual label and the automatic result can still be acceptable. Previously, Renna \textit{et al.} \cite{Renna-UNET} used a delay tolerance-based definition, where the problem was transformed into a classification problem, thus common accuracy measures could be used. The authors described a detection as true positive (TP) if it was inside the tolerance interval around a manual label, else it was deemed as a false positive (FP). Since negative classifications were not described, not all accuracy measures could be applied. In the same publication, the authors used F1-score and positive predictive value (PPV). Based on their work we used the same definitions and the same performance metrics which were
\begin{equation}
    PPV = \frac{TP}{TP+FP},
\end{equation}
\begin{equation}
    F1 = \frac{2 \times PPV \times TPR}{PPV + TPR},
\end{equation}
where $TPR$ is the true positive rate or sensitivity, which can be calculated from the ratio of TP cases and all positive cases:
\begin{equation}
    TPR = \frac{TP}{M},
\end{equation}
where $M$ is the number of manual labels.
The previously mentioned tolerance value can be changed dynamically and these performance values will change accordingly. We introduced a method so that this dependence can be graphed similar to a receiver operating characteristic (ROC) curve. This is called a \textit{Score-vs-Tolerance} plot.  In our case, the minimal tolerance was set to 5 ms and the maximal tolerance was 90 ms. By setting a threshold for the performance score, we can determine at which tolerance value can we expect the given performance score. This can be annotated on the \textit{Score-vs-Tolerance} plot as well.
Additionally, we introduced the mean of absolute differences (MAE) of the detections and the manual labels, defined as
\begin{equation}
    MAE = \frac1N \sum_{k=0}^N \left| d_k-l_k \right|,
\end{equation}
where $N$ is the number of detections, $d_{k}$ is the k-th detection location and $l_{k}$ is the closest manual label to $d_{k}$. With this score, a smaller value is considered better. To unify the above scores to all test signals we accumulate the TP and FP values in a sum, and the MAE scores in a list. PPV and F1 were calculated with these sums, and the final MAE was acquired by taking their mean and standard deviation.

For the naive detection method the preset minimal distance in our case was 270 ms. This represents a maximal 222 beats per minute (BPM) heart rate (60/beat-to-beat time in seconds), which is well above the maximal expected heart rate even for a fetus (110-180 BPM). It is important to note that these methods usually detect the largest local maxima which satisfy the prescribed properties. In our case this means that the detected peaks will correspond to the louder heart sounds, in most cases this means the detection of the S\textsubscript{1} sounds only. Since our emphasis was placed on accurate S\textsubscript{1} detection this is acceptable, however this means that further steps are needed to detect S\textsubscript{2} sounds, which were not included. The Neurokit peak detection method is similar to the naive method but with the main difference, that the minimal distance was not configured, however, a lower threshold for the peaks was set as the median of the homomorphic envelope.

For both the \textit{HSMM} and \textit{pyHSMM} models, they were trained with the same dataset as they were benchmarked on but with every fifth signal removed to prevent training biases. The state segments were converted to detection locations by taking the center of each segment.

\section{Results}
\begin{figure}
    \centering
    \includegraphics[width=0.8\textwidth]{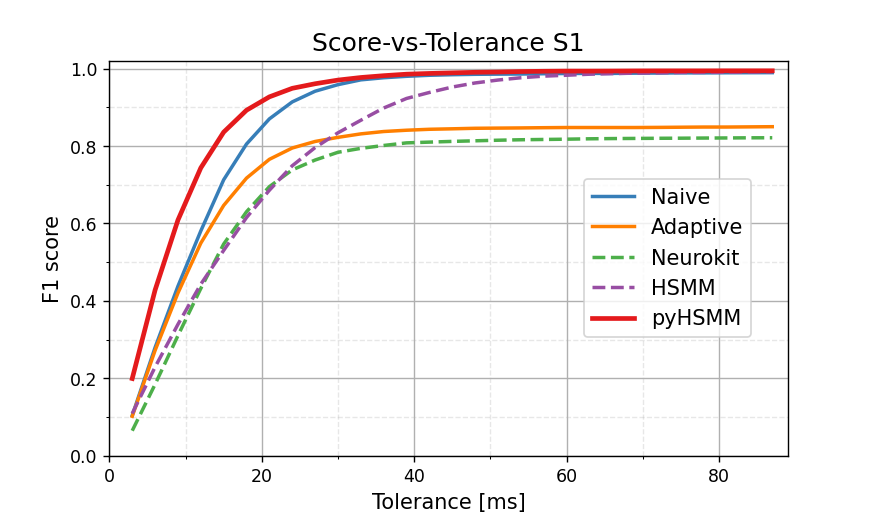}
    \caption{Unified Score-vs-Tolerance plot for the tested methods detecting S\textsubscript{1} sounds.}
    \label{fig:acc-s1}
\end{figure}

\begin{figure}
    \centering
    \includegraphics[width=0.8\textwidth]{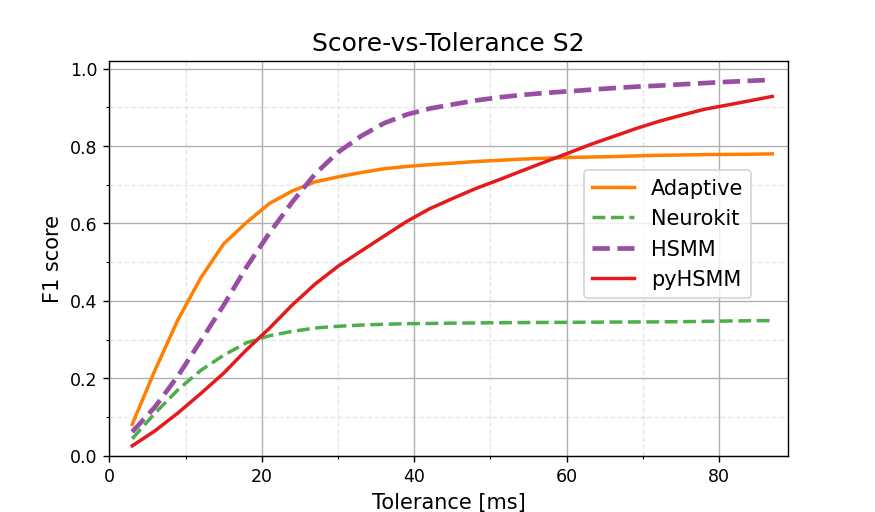}
    \caption{Unified Score-vs-Tolerance plot for the tested methods detecting S\textsubscript{2} sounds.}
    \label{fig:acc-s2}
\end{figure}

The previously mentioned performance metrics (PPV, F1, MAE) were calculated for each detection method. For PPV and F1-score a constant tolerance of 30 ms was chosen based on the nature of the data. This benchmarking was performed on both S\textsubscript{1} and S\textsubscript{2} sounds. In applicable cases, the Score-vs-Tolerance curves were also calculated with the F1-score and plotted. The performance scores can be seen in \Tref{tab:accs} for S\textsubscript{1} and \Tref{tab:accs-S2} for S\textsubscript{2} sounds. While the Score-vs-Tolerance plots are shown in \Fref{fig:acc-s1} for S\textsubscript{1} and \Fref{fig:acc-s2} for S\textsubscript{2}.

\begin{table}
\setlength{\tabcolsep}{15pt}
\caption{Performance metrics of the tested methods for S\textsubscript{1}}
\begin{center}
\begin{minipage}{300pt}
\lineup
\begin{tabular}{@{}llll}
\br
Method   & PPV (\%) & F1 (\%) & MAE (ms)        \cr \mr
Naive    & 96.1     & 95.9    & 14.3 $\pm$ \08.5    \cr
Adaptive & 79.9     & 82.3    & 33.1 $\pm$ 25.2   \cr
Neurokit & 95.0     & 78.4    & 61.7 $\pm$ 27.9   \cr
HSMM     & 83.5     & 83.4    & 19.2 $\pm$ 10.4   \cr
pyHSMM   & \textbf{97.0}     & \textbf{97.1}   & \textbf{10.8 $\pm$ 7.9}    \cr
\br
\end{tabular}
\fontsize{8pt}{10.5pt}
\selectfont
\textsuperscript{*}Best results highlighted in bold
\label{tab:accs}
\end{minipage}
\end{center}
\end{table}

\begin{table}
\setlength{\tabcolsep}{15pt}
\caption{Performance metrics of the tested methods for S\textsubscript{2}}
\begin{center}
\begin{minipage}{300pt}
\lineup
\begin{tabular}{@{}llll}
\br
Method   & PPV (\%) & F1 (\%) & MAE (ms)       \cr \mr
Naive    & N/A      & N/A     & \0N/A            \cr
Adaptive & 68.8     & 72.0    & \047.9 $\pm$ 44.1  \cr
Neurokit & 40.6     & 33.4    & 131.7 $\pm$ 27.8 \cr
HSMM     & \textbf{78.5}     & \textbf{78.4}    & \textbf{25.6 $\pm$ 16.7}  \cr
pyHSMM   & 48.8     & 48.9    & \040.5 $\pm$ 22.3  \cr
\br
\end{tabular}
\fontsize{8pt}{10.5pt}
\selectfont
\textsuperscript{*}Best results highlighted in bold
\label{tab:accs-S2}
\end{minipage}
\end{center}
\end{table}

The pyHSMM method had the best results for S\textsubscript{1} detection. The score-vs-tolerance curve show that the minimum tolerance to a 0.8 F1 score threshold was only around 13.50 ms, and the score plateaued quickly as the tolerance increased. The other detection algorithms also performed well, however, all of them had their drawbacks. The naive method was accurate but it can only detect S\textsubscript{1} sounds, Neurokit could detect the sounds accurately but a large amount of false positives were introduced, HSMM performed lower than expected which is most likely due to the lack of proper fine-tuning, finally the adaptive method had the lowest accuracy which is presumably caused by the S\textsubscript{1}-S\textsubscript{2} separation.

Achieving good results for S\textsubscript{2} detection are currently a lower priority for us because these can be improved by taking into consideration S\textsubscript{1} detections, however, they can be used to contrast the performance of the tested methods. For S\textsubscript{2} the HSMM method was the best in all regards, both in the numerical results and in the Score-vs-Tolerance result. All the other methods had a significant drop in accuracy.

\begin{figure}
    \centering
    \includegraphics[width=0.49\columnwidth]{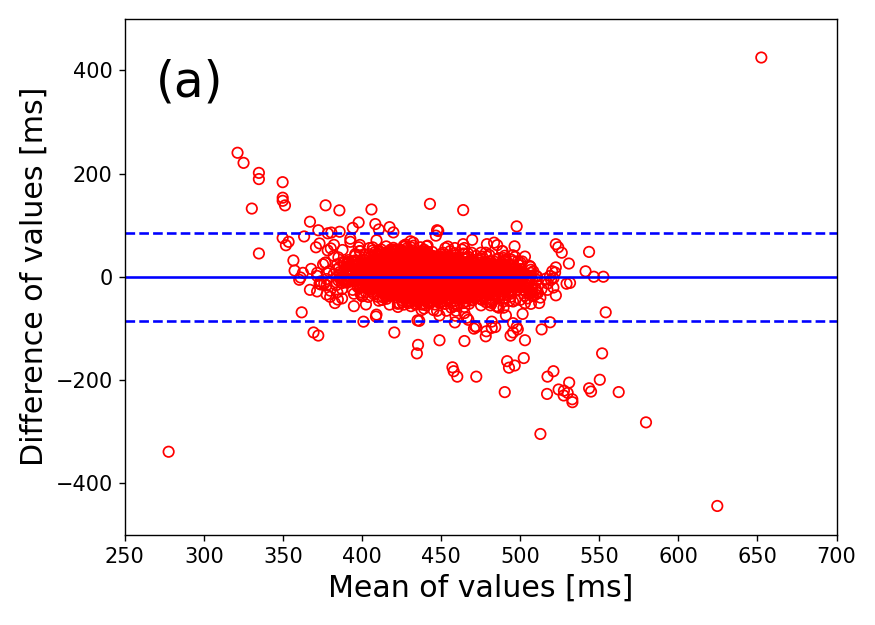}
    \includegraphics[width=0.49\columnwidth]{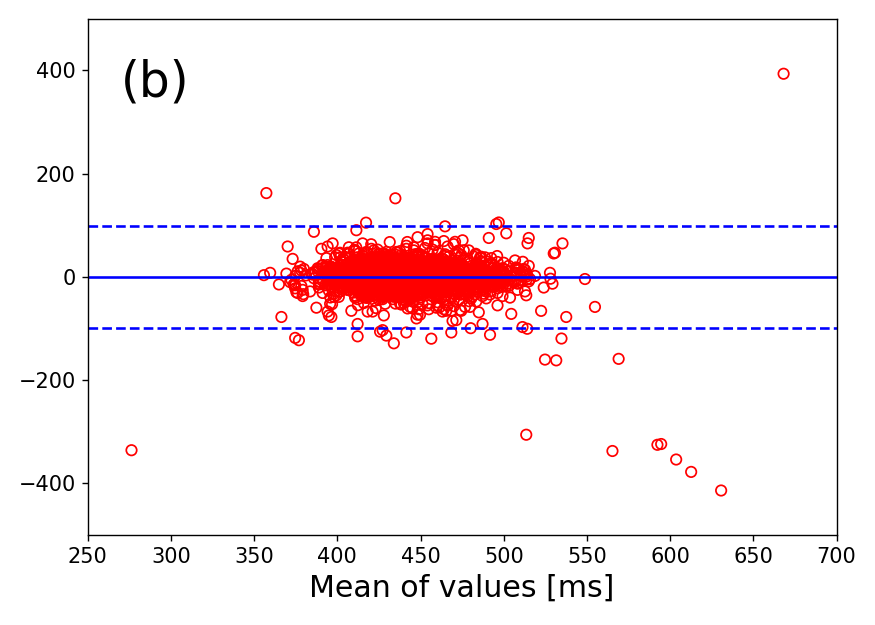}
    \caption{Bland-Altman plot for S\textsubscript{1} detection (a): HSMM method (2 outliers removed), (b): pyHSMM method (8 outliers removed), the solid blue line gives the mean of differences and the dashed blue line marks the 1.96*std interval.}
    \label{fig:BA-S1}
\end{figure}

\begin{figure}
    \centering
    \includegraphics[width=0.49\columnwidth]{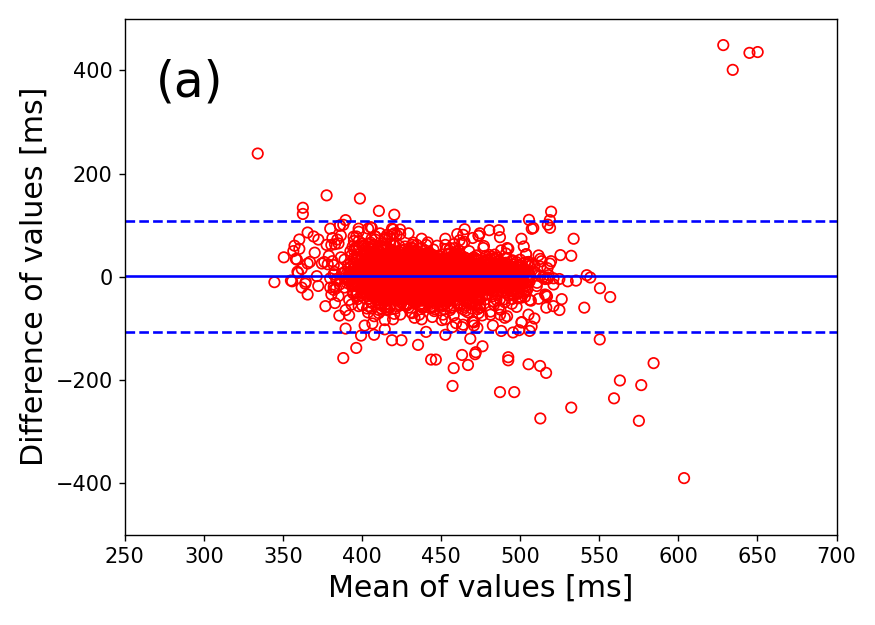}
    \includegraphics[width=0.49\columnwidth]{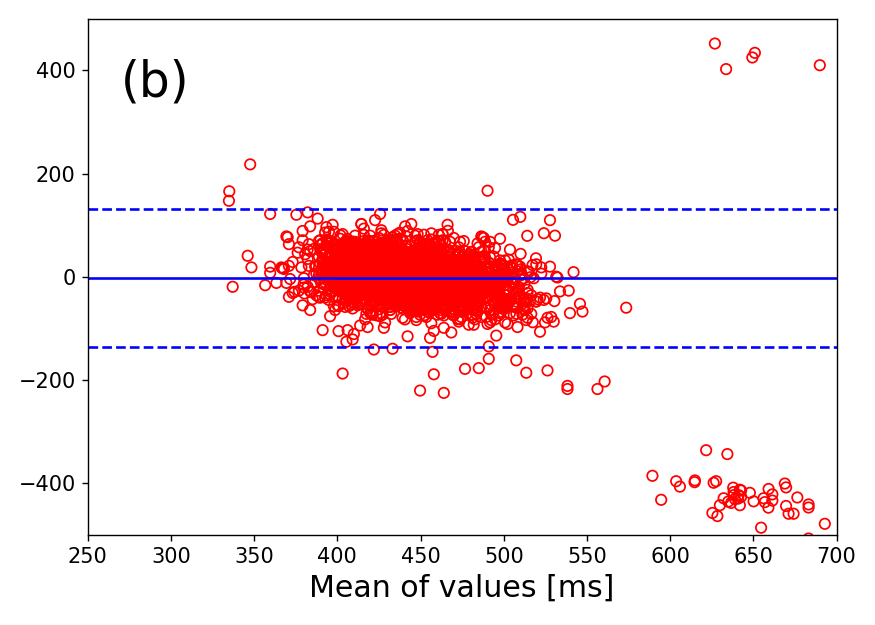}
    \caption{Bland-Altman plot for S\textsubscript{2} detection (a): HSMM method (7 outliers removed), (b): pyHSMM method (15 outliers removed), the solid blue line gives the mean of differences and the dashed blue line marks the 1.96*std interval.}
    \label{fig:BA-S2}
\end{figure}

For further analysis the results of the pyHSMM and HSMM methods were compared with the ground truth manual labels in Bland-Altman plots. To remove the time dependence of the labels, the consecutive differences of the labels were compared instead of the raw values, similarly to \cite{Leikan2016-baplot}. Then the means and the difference between the ground truth and the detections were plotted. The results for S\textsubscript{1} and S\textsubscript{2} detections were plotted in different figures, see \Fref{fig:BA-S1} and \Fref{fig:BA-S2}.


\section{Discussion}
Our toolbox implements most widely used processing methods for PCG analysis, with the possibility of future extension with additional features and more sophisticated segmentation models. The design of our proposed framework is highly modular and the building blocks for a given processing flow are easily interchangeable. Each step can be evaluated and tested individually, but these steps can be also chained into more complex processes. To simplify this we provided several tools to reduce code repetition and improve future parallelizability. In our opinion this proposed toolbox achieves the goals which were outlined in Section \ref{ssec:motivation} and can be used to process both fetal and non-fetal PCG.

Although the chosen SQI features in \ref{ssec:sqi} were originally designed to be used as inputs to a support vector machine (SVM) which would learn to differentiate between the quality classes, we included them without the SVM as just raw values to determine the quality of the signal. This way the user has to decide whether to accept a signal, reject it, or to apply further processing.

Our pyHSMM detection performed significantly worse than expected for S\textsubscript{2} segmentation. The most probable cause for this is inaccurate systole length estimation, since this is the most significant modification we have made the original Springer implementation. This was done to better capture the properties of the fetal heart cycle, although this results suggest that a more accurate method is required. Although as mentioned before S\textsubscript{2} detection is a lower priority, since with accurate S\textsubscript{1} detection the accuracy of S\textsubscript{2} can be improved with a multi-stage process. For example with detecting peaks in the envelope between the S\textsubscript{1} sections.

\ack
The dataset used to build the validation data was provided by Prof. Ferenc Kovács.

Project no. TKP2021-NVA-27 has been implemented with the support provided by the Ministry of Culture and Innovation of Hungary from the National Research, Development and Innovation Fund, financed under the TKP2021 funding scheme.

\section*{Code availability}
The latest version of pyPCG can be downloaded from its Github repository:\\ \url{https://github.com/mulkr/pyPCG-toolbox}.\\For the documentation see:\\ \url{https://pypcg-toolbox.readthedocs.io/en/latest/}.\\
Segmentation validation code available at:\\ \url{https://colab.research.google.com/drive/1TQrgCDys47YJ_bu-iYo_HcB3LF03ixVv?usp=sharing}\\
Validation data is available upon request to the authors.

\section*{References}
\bibliography{refs}
\bibliographystyle{iopart-num}
\end{document}